# Thermal effects of late accretion to the crust and mantle of Mercury

**S.J. Mojzsis** [1,2,†], **O. Abramov**[3†], **E. A. Frank**[4,5], **and R. Brasser**[6†]

[1] Department of Geological Sciences, University of Colorado, UCB 399, 2200 Colorado Avenue, Boulder, CO 80309, USA.

[2] Institute for Geological and Geochemical Research, Research Center for Astronomy and Earth Sciences, Hungarian Academy of Sciences, 45 Budaörsi Street, H-1112 Budapest, Hungary.

[3] Planetary Science Institute, 1700 East Fort Lowell Road, Suite 106, Tucson, AZ 85719, USA.

[4] Department of Terrestrial Magnetism, Carnegie Institution of Washington, 5241 Broad Branch Rd, Washington, DC 20015, USA.

[5] Planetary Resources, 6742 185th Ave, NE Redmond, WA 98052, USA.

[6] Earth Life Science Institute, Tokyo Institute of Technology, Ookayama, Meguro-ku, Tokyo 152-8550, Japan.

\* Corresponding author: Stephen J. Mojzsis  email: mojzsis@colorado.edu

† Collaborative for Research in Origins (CRiO), the John Templeton Foundation – FfAME Origins Program.

**Highlights:**

- Mercury's silicate reservoirs were modified by impact bombardment during late accretion (ca. 4.4-3.8 Ga).

- Late accretion bombardment melted Mercury's crust and enhanced its mantle's thermal profile.

- Impactors ≳100 km in diameter could produce high-temperature, ultramafic melts and effusive volcanism to form Mercury's observed high-magnesium region (HMR).

**Abstract**

Impact bombardment on Mercury in the solar system's late accretion phase (ca. 4.4-3.8 Ga) caused considerable mechanical, chemical and thermal reworking of its silicate reservoirs (crust and mantle). Depending on the frequency, size and velocity of such impactors, effects included regional- and global-scale crustal melting, and thermal perturbations of the mercurian mantle. We use a 3D transient heating model to test the effects of two bombardment scenarios on early (pre-Tolstojan) Mercury's mantle and crust. Results show that rare impacts by the largest ($\gtrsim$100 km diameter) bodies deliver sufficient heat to the shallow mercurian mantle producing high-temperature ultra-magnesian (komatiitic *s.s.*) melts. Impact heating leading to effusive (flood) volcanism can account for eponymous "High-Magnesium Region" (HMR) observed during the MErcury Surface, Space Environment, GEochemistry Ranging (MESSENGER) mission. We find that late accretion to Mercury induced volumetrically significant crustal melting ($\leq$58 vol.%), mantle heating and melt production, which, combined with extensive resurfacing ($\leq$100%), also explains why its oldest cratering record was effectively erased, consistent with crater-counting statistics. (**164 words**)

**1 Introduction**

The first (remote) geochemical measurements made by the X-Ray Spectrometer (XRS) on the MErcury Surface, Space Environment, GEochemistry, and Ranging (MESSENGER) spacecraft, coupled with geomorphological evidence, suggested the presence of ancient (>3.9 Ga corresponding to Pre-Tolstojan; Marchi et al., 2011; Werner, 2104) komatiite-like crust on Mercury (Head et al., 2011; Nitter et al., 2011). Komatiite is an extrusive magnesium-rich (MgO

≥18 wt.%) rock that on Earth is most often found associated with Proterozoic (2500 – 0.542 Ma) and older terranes extending to the Eoarchean (Arndt et al., 2008; Frank et al., 2016). Curiously, there are also rare occurrences of Phanerozoic komatiite-like basalts related with (super-)plume activity such as at the late Mesozoic (ca. 90 Ma) Gorgona Island locality in Columbia (Aitken & Echeverría 1984), and the Permo-Triassic (ca. 270 Ma) Song Da komatiites in Vietnam (Hanski et al. 2004). It is the usual case, however, that komatiitic lavas are otherwise restricted to ancient rocks in the terrestrial record. This observation, coupled with petrological arguments and field observations, has been used to argue that komatiitic lavas were commonplace on the Archean Earth owing to the fact that mantle temperatures were higher than present (by 200º C or more; Arndt et al., 1997). Yet, as with many other aspects of the "early Earth" record, the petrologic and geodynamic conditions for the origin and seemingly widespread emplacement of komatiites in the first half of Earth's history are disputed (Grove et al., 1999). The general, but not unanimous, consensus is that komatiite genesis on Earth is due to ultra-high temperature mantle plumes that ascend a deep thermal boundary layer, possibly at the core-mantle boundary (CMB; Arndt, 2003).

These ostensibly rare and ancient ultramafic lavas also appear to exist on other solid bodies in the solar system: komatiites have been proposed as an analogue for high Mg- and Fe-basalts from Gusev Crater on Mars (e.g. Bost et al. 2012), as erupting from contemporary volcanoes on Jupiter's moon, Io (Williams et al. 2000), and as an explanation for the High-Magnesium Region (HMR) on Mercury (e.g. Weider et al., 2012). A lunar komatiite component was insinuated from compositional end-member mixing models of an Apollo 16 highland breccia (Ringwood et al., 1986). Setting aside Mars for the time being, the existence of komatiites on the Moon (cf.

Huppert and Sparks, 1985) and Mercury would be surprising; these are the smallest and most thermally moribund worlds of the inner solar system, and if komatiites are the result of high degrees of partial melt from deep hot plumes, as has been invoked for terrestrial komatiites stated above, it is difficult to reconcile this mechanism with what we understand about the interior structure of either. Mercury's anomalously high density (5.43 g cm$^{-3}$) as well as other geophysical data, suggests a very shallow CMB depth of only ~420 km (Hauck et al. 2013) and thus a core that comprises ~60% of its volume with a correspondingly small mantle (Smith et al. 2012). The Moon is nearly the geophysical opposite of Mercury. It has an unusually low density (3.35 g cm$^{-3}$; approaching that of Earth's upper mantle) and a core that is just about 20% of its volume; most of the Moon's volume is occupied by its mantle.

The initial observations of komatiite-type crustal regions on Mercury were upheld by subsequent higher-resolution XRS and Gamma Ray and Neutron Spectrometer (GRNS) measurements and further petrologic models (Charlier et al., 2013; Peplowski et al., 2015; Stockstill-Cahill et al., 2012; Weider et al., 2012, 2015; Vander Kaaden et al., 2017). One geochemical terrane was shown to be especially magnesium-rich, the eponymous "high-Mg region" (HMR; Weider et al., 2015). The HMR is defined by a large, contiguous crustal domain with [Mg/Si] >0.5; it covers ~15% of Mercury's surface, yet is not associated with any geological or geophysical features that could point to a clear origin whether by impact or effusive (rift?) volcanism (Frank et al., 2017). Volcanic eruptions could be facilitated by impacts (Green, 1972; Williams and Greeley, 1994), but are not necessarily initiated by them (Ivanov and Melosh, 2003; cf. Reese and Solomatov, 2006). Systematic searches through MESSENGER datasets for evidence linking the HMR to an ancient (and cryptic) impact basin led Frank et al. (2017) to conclude that it is more likely the product of rift volcanism from a chemically and thermally heterogeneous mercurian mantle, rather than a direct consequence of impact-excavated mantle material.

In this work, we investigate the thermal and mechanical consequences of impact bombardment to Mercury's crust and mantle to understand the HMR, and explore the possibility that ultramafic (ultra-magnesian) melts can be generated by altering the thermal structure of the mercurian mantle. Our work is thus complementary to Rolf et al. (2017) who explored how impacts affected the evolution of the Moon's internal thermal state ~100 Myr after bombardment.

For clarity and consistency with other work, we define the interval of "late accretion" in the solar system as the time subsequent to the separation of the silicate reservoirs of the sampled inner planets (Earth and Mars) after approximately 4.5 Ga (e.g Brasser et al., 2016 and references therein). The late accretion interval thus corresponds to impact processes which occurred well after the conclusion of primary accretion associated with the bulk of planet formation. Such late additions nevertheless induced profound chemical and mechanical modifications of solid (silicate or icy) planetary crusts for hundreds of millions of years (Melosh, 1989). Abundant lines of evidence, based primarily on lunar studies, show that shock-metamorphism – including small degrees of melting – extended into the Archean and Paleoproterozoic eons (reviewed in Bottke and Norman, 2017). Samples from the Moon show evidence for relatively late impact reset ages at ca. 3.95-3.85 Ga that correspond to estimates for the age of Imbrium basin (e.g., Hopkins and Mojzsis, 2015 and references therein). The existence and nature of a late heavy bombardment (LHB) is, however, debated (cf. Boehnke and Harrison, 2016). A common interpretation of the record of late accretion is that of a superimposed LHB impactor uptick (Ryder, 1990) from what is almost certainly an overprint of the Imbrium basin-forming event (Bottke and Normam 2017) on an otherwise extended monotonic decline of the impactor flux to the inner solar system (Dalrymple and Ryder, 1993, 1996; Tera et al., 1974; Turner et al., 1973). The duration of a proposed LHB spike, even if it is real, is also unclear and controversial, but may have lasted between 20 – 200 Myr (cf. Hartmann et al., 2000). Meteorites derived from multiple parent bodies in the asteroid belt, as well as the ancient martian meteorite ALH84001, further show effects of impact-induced metamorphism with some rare dates that overlap that of Imbrium cited above (e.g., Bogard, 1995; Ash et al., 1996; Kring and Cohen, 2002; Swindle et al., 2009; Marchi et al., 2013; Hopkins et al., 2015). Many have used this observation to shore-up the argument for a solar system-wide LHB. Lastly, the contemporary population structure of the main asteroid belt can be explained by late-stage giant planet migration (Minton and Malhotra, 2009) as described in the popular Nice model (Tsiganis et al., 2005); the Nice model outcome has been widely invoked to explain the LHB (Gomes et al., 2005). Yet it is important to note that the absolute timing of this postulated late-stage migration is loosely constrained, with a dynamic uncertainty of at least 250 Myr (e.g., Morbidelli et al., 2012). In aggregate, although data suggest that bombardment between ca. 4.5 – 3.8 Ga affected the inner solar system by resurfacing or even melting substantial fractions of the terrestrial planets' crusts (Tonks and Melosh, 1993), the precise timing and intensity of this bombardment remain unclear.

We examine whether impact bombardment to Mercury's shallow mantle in its pre-Tolstojan eon could be a viable mechanism for triggering the production of komatiite-like melts at the regional-scale (cf. Frank et al., 2017). To do this we assess the increased impact flux during late accretion in two bombardment scenarios, roughly coincident with strong surface modification on Mercury (Marchi et al., 2013; Werner, 2014) and determine whether this mechanism could have induced sufficient temperature increase in the mantle – assuming peridotite compositions – to enable crossing of the komatiite solidus. Explaining the genesis of such high-temperature ultra-high magnesian silicate melts contributes to our assessment of the degree of crustal melting due to sustained periods of bombardment across the inner solar system, and how such resurfacing can reset crater counting statistics.

## 2. Model description

Considering the uncertainties in both the timing and duration of late accretion to Mercury, we analyze two bombardment scenarios: The first is the "classical Late Heavy Bombardment," (cLHB) as described in Abramov and Mojzsis (2016); second is a hybrid-LHB characterized by a brief increase in impacts followed by an extended decay over about 400 Myr ("sawtooth Late Heavy Bombardment," sLHB; Morbidelli et al., 2012; Abramov and Mojzsis, 2016; cf. Turner et al., 1979). We refer the reader to Figure 1 of Abramov and Mojzsis (2016) for a representation of the time-flux evolution of these two models. Recent successors to the Nice model (Morbidelli et al., 2012; Marchi et al., 2012, 2014; reviewed in Bottke and Norman, 2017) favor the sawtooth scenario, but this too is debated (e.g., Kaib and Chambers, 2016). We reserve our models to the two general cases described above, and we assess the viability of the bombardment mechanism to heat the planet's mantle sufficiently to generate komatiite-type melts and to both thermally and mechanically modify its crust.

Using the Marchi et al. (2013) impact flux parameters for Mercury, we simulate late accretion in both a cLHB and sLHB model, following Abramov and Mojzsis (2016). It is worth mentioning that in that study, which was applied to Mars, we modeled a cLHB scenario as a spike in impactors at ca. 3900 Ma and constant flux over the subsequent 100 Myr (e.g. Abramov and Mojzsis, 2009). The cLHB scenario can also be viewed as broadly representative of the Neukum and Ivanov (1994) "classic post-accretionary" timeline between ca. 4400 – 4300 Ma (or between about 4500 – 4400 Ma on the sawtooth timeline of Morbidelli et al., 2012). The sLHB scenario is defined as a continuation of the Neukum and Ivanov (1994) "classic post-accretionary" timeline extending into the earliest vestiges of Earth's geological record in the late Hadean (ca. 4200 – 3850 Ma) and Archean eons (ca. 3850 – 2500 Ma; cf. Bottke et al., 2012, Morbidelli et al., 2012). Total mass delivered in our sLHB model is a factor of ~3 less than that in the cLHB. As in Abramov and Mojzsis (2016), we also analyze boundary cases of the cLHB and sLHB, each with a factor of 10 greater delivered mass than its respective baseline case. We do this to address regional temperature changes in the mantle arising from very large (>200 km diameter) impactors, and the cumulative effect of higher delivered mass on global crustal melting.

To track the instantaneous and cumulative temperature changes due to impacts as a function of position and depth across the planet's surface, crust and mantle, we apply to pre-Tolstojan Mercury a modified version of the global cratering model previously used in Abramov and Mojzsis (2009) and Abramov et al. (2013) for the Hadean Earth, and in Abramov and Mojzsis (2016) for pre-Noachian Mars. Detailed descriptions of these models can be found in Abramov et al. (2013).

We now summarize the main features to distinguish applications unique to this analysis of early Mercury: Model inputs including the location, diameter, and velocity of each impact are generated by a stochastic cratering model (Richardson et al., 2005), which is used to populate all or part of a planetary surface with craters as a function of time within constraints established from both models and observations (see Section 2.1). For each crater in the model a temperature field is calculated using analytic expressions for shock deposited heat and central uplift, as described in Abramov et al. (2013). After the crater's thermal field is introduced into a three-dimensional model representing the planet's crust and mantle, it can cool by conduction in the subsurface and by radiation at the surface. The analytic method used to calculate post-impact temperature distributions has been shown to agree well with hydrocode modeling results (Abramov and Mojzsis, 2016).

The crust and mantle are modeled as a cuboid, such that the surface area in the model is constant with depth. Mercury's total volume is a factor of 1.7 that of its core, with the volume of our model at the CMB thus oversized by this factor. We analyze temperature changes in localized regions so that, at most, our analyses probe a few percent of the model surface area at a given depth. As such, quoted temperature increases and komatiite volumes are reasonable estimates (and lower bounds on what would be obtained with a spherical model geometry). Internal temperatures are recorded every $10^5$ years from the surface to a depth of 420 km, which we take as the CMB (see Section 2.1.4). Approximately 5,800 model elements of area $1.3 \times 10^4$ km$^2$ are monitored at each depth, with a separation between depths of 10 km. This allows us to track the localized temperature effect of each impactor across depth as well as the lateral and vertical heat transfer between elements. Williams et al. (2011; see also Rivera-Valentin and Barr, 2014) described both latitude and longitude temperature variations in Mercury's lithosphere. This is not especially relevant for our focus on *strictly regional (local)* temperature increases as opposed to global-scale effects.

## 2.1 Input parameters

### 2.1.1 Delivered mass

We use a Mercury/Moon impact flux ratio of 1.5 (Marchi et al., 2009), expressed as impacts per unit area per unit time over the period of late accretion as defined herein. Mercury's surface area is about twice that of the Moon, which yields ~3 impacts on Mercury for every lunar impact with this flux ratio. Abramov et al. (2013) estimated $2 \times 10^{20}$ kg total mass delivered to the Earth under the cLHB scenario (e.g., Levison et al., 2001), corresponding to ~$10^{19}$ kg for the Moon (e.g., Morbidelli et al., 2012), and $3 \times 10^{19}$ kg for Mercury according the the above relation. Similarly, Abramov and Mojzsis (2016) found $3 \times 10^{19}$ kg delivered to Mars under the sLHB scenario, which corresponds to $3 \times 10^{18}$ kg for the Moon and $8 \times 10^{18}$ kg for Mercury. The total mass in each scenario is fixed to the size-frequency distribution (SFD) described in the next section to produce the distribution of impacts over time. We note that these mass estimates are not well constrained at the high-end due to effective crater saturation and erasure (e.g., Richardson, 2009) and that a factor 10 increase in delivered mass can significantly affect results by increasing the probability of extremely large (≳100 km diameter) impactors (see Supplementary Online Materials). This uncertainty motivates our inclusion of 10x delivered mass in the extreme cLHB and sLHB scenarios.

### 2.1.2 Impactor parameters

Late accretion impactors may have been dominated by a population whose size-frequency distribution was roughly similar to the present-day main belt asteroids (Strom et al. 2005), if this analysis is correct then the SFD has not changed significantly since the late accretionary bombardment interval as we define it (Bottke et al., 2005). In our model, we use the current main belt SFD and consider only impactors >5 km in diameter to analyze the thermal fields of thousands rather than millions of impact events to keep computation time within reason. As pointed out in our previous work (Abramov and Mojzsis, 2016), a high-resolution, static model that includes smaller impactors shows differences from the type of dynamic model used in this work at the ≤20% level in cumulative crustal heating. We thus take our quoted values for cumulative crustal heating as reasonable minimum estimates. Moreover, along with this conservative approach we also use a moderate mean impact velocity estimate for Mercury of 33 km s$^{-1}$, compared to higher estimates of up to about 43 km s$^{-1}$ (LeFeuvre and Wieczorek, 2008). With the energy delivered by an impactor scaling as the square of its velocity, the difference between the two velocities is therefore substantial. Accordingly, our predictions of komatiite-type melts produced by partial melting in the mantle due to impact heating represent conservative values, which is appropriate when seeking to investigate the feasibility of bombardment as a mantle heating mechanism. In all model outputs reported here, we assume bombardment by rocky asteroids with a mean density of 2800 kg m$^{-3}$ and the most probable impact angle of 45° (e.g., Shoemaker, 1962).

**2.1.3 Crustal target parameters**

Rock density, thermal conductivity and heat capacity of the target are assumed to be those of basalt, the dominant lithotype of Mercury's crust (Stockstill-Cahill et al., 2012; Vander Kaaden et al., 2017) and the expected partial melt product of a mantle of peridotitic composition. We take the crustal thickness to be 35 km (Padovan et al., 2015). Pressure derivative of the bulk modulus is 5.5, and adiabatic bulk modulus at zero pressure is $19.3 \times 10^9$ Pa (Gault and Heitowit, 1963). Input target properties are summarized in **Table 1**.

**2.1.4 Mantle boundary depths, adiabatic gradient and hermeotherm**

We assume an average surface temperature of 120 °C using a solar luminosity at 3.9 Ga of 0.7 $L_\odot$ (Sackmann and Boothroyd, 2003) and assuming Mercury was in its current orbit; this may not have been true prior to the LHB (Brasser et al., 2013). The day-night difference in surface temperatures on contemporary Mercury are large ($\Delta T_{surface}$ ~600K), and because we neglect this feature the different initial surface temperatures as a function of solar evolution or orbital changes are unimportant. The CMB is set at a depth of 420 km based on estimates of Smith et al. (2012) and Hauck et al. (2013). Michel et al. (2013) and Tosi et al. (2013) approximate the base of the lithosphere at roughly at 1/4 – 1/3 the combined crust and mantle thickness. We take this boundary to be at a depth of 120 km, or about 30% the combined thickness. A reasonable temperature range for the lithosphere base is 1475 – 1675 °C (Namur et al., 2016a,b; Vander Kaaden and McCubbin, 2015), with high uncertainty on both ends. We use a temperature range at the CMB of 1475 – 1825 °C as in Hauck et al. (2013). The extreme ends of these temperature ranges suggest a thermal gradient (the equivalent geotherm on Mercury is hereafter referred to as the *hermeotherm*) of 0 – 1.2 °C km$^{-1}$ between 120 – 420 km depth which is the distance between the lithosphere base and the CMB.

We note the starting absolute temperature profile in the mantle, and the resulting (deep) thermal gradient, strongly influences the amount of komatiite production. A steep hermeotherm would seem to mitigate much mantle convection on either present or past Mercury under reasonable viscosity regimes and with a thick thermal boundary layer (e.g., Redmond and King, 2007; Ogawa, 2016). We thus consider two cases for the temperatures at the lithosphere base and CMB, with accompanying hermeotherm: (i) 1575 – 1650 °C from the lithosphere base to the CMB (an average of each temperature range above), which gives an approximate 0.3 °C km$^{-1}$ hermeotherm; and (ii) 1475 – 1825 °C (the maximum temperature range of the above estimates), with 1.2 °C km$^{-1}$. In analyzing crustal melts, we assume a hermeotherm of about 12 °C km$^{-1}$ in the lithosphere, derived from estimates of average surface temperature (120 °C; a relatively insensitive parameter in our model), lithospheric base temperature (approximately 1575 °C), and lithospheric base depth (120 km) as discussed above.

**2.1.5 Conditions for komatiite formation on Mercury**

Komatiite magmas form either by enhanced temperatures which induce very high degrees of partial melting (≥50%) and/or by melting at high pressure (>15 GPa; Green, 1975; Robin-Popieul et al., 2012). For Mercury, Hauck et al. (2013) predicted pressures of only about 5.7±0.5 GPa at 420 km depth; we place the CMB at this depth, thus excluding the high-pressure formation mechanism. High degrees of partial melting due to elevated temperatures, however, remain plausible. On Earth, there is experimental and observational evidence that komatiite formation requires temperatures of at least 1750 °C under dry conditions at pressures of approximately 6 GPa (Herzburg and Zhang, 1996; Robin-Popieul et al., 2012). Applying these assumptions to our temperature case (ii) described in Section 2.1.4, with a CMB temperature of 1825 °C, would suggest komatiite generation near the mercurian CMB without any need for impact-induced heating. We may then expect these melts would be delivered to the surface via plumes, and that HMRs would cover much more of Mercury's surface than what is observed by MESSENGER. Instead, to assess the viability of komatiite production by impact-induced mantle heating, we perform our analysis under the more conservative case (i), in which ambient mantle temperatures and pressures are too low to generate komatiitic melts (even if decompression melting is taken into account; Grove and Parman, 2004). In this elementary case, we use the phase diagrams in Robin-Popieul et al. (2012) to determine when and where a temperature increase of about 150 °C generates the required partial melting of >50% to produce komatiite throughout Mercury's asthenosphere. Our computationally simple approach was that for each model element in the asthenosphere (120 – 420 km depth) at each timestep, we consider the change from initial temperature at the simulation start ($\Delta T$). For timesteps in which $\Delta T \geq 150$ °C at a given element, komatiite melt is taken to form within that element's volume. The number of model elements at a given timestep that meet this criterion yields the volume of the asthenosphere over which komatiite is produced.

## 3. Results and analysis

Results of each model are summarized in **Table 2**. Reported values are most sensitive to our assumptions for: (i) total delivered mass – in addition to being proportional to delivered energy, increasing delivered mass by a factor of 10 can be approximated by doubling the diameters of the largest impactors; this case would deliver a factor of ~8 greater energy per impact to twice the depth (Pierazzo et al., 1997; Pierazzo and Melosh, 2000); (ii) the primordial asteroid belt size-frequency distribution. The SFD we use favors larger impactors, which results in delivering proportionally more energy into the mantle; and (iii) impactor velocity, when squared, is proportional to delivered energy. Important for this study, the stochastic time and position of impacts can also have strong temperature consequences in localized regions, e.g., when large impacts are separated by relatively short time and distance. For example, in the cLHB case, while the largest instantaneous increase at the CMB is about 400 °C due to a single 200 km diameter impactor, the largest cumulative temperature increase ($\Delta T$) is 672 °C from a 100 km diameter impactor subsequently striking the surface within 500 km, resulting in a significant mechanical and thermal overlap. It is impacts by bodies ≳100 km in diameter that raise temperatures in localized regions of the asthenosphere to (and well in excess of) our threshold for komatiite production described in Section 2.1.5. Basin-forming events are required to produce ultra-magnesian melts in Mercury's mantle, regardless of the bombardment model used.

The model output in the baseline cLHB case is shown in **Figure 1 (a) – (c)** for the timestep following the largest (200 km diameter) impactor in our SFD, 10 Myr later, and the final timestep. **Figure 1 (d) – (e)** show the sLHB model following the largest impactor and at the final timestep. The paucity of large impactors in the sLHB relative to the cLHB scenario is apparent in the comparatively weak cratering and largely unperturbed asthenosphere temperature profile in the former. **Figure 1 (f)** shows the temperature effect of the cLHB model's largest impactor (as well as all others) over time and as a function of depth, demonstrating the long cooling timescale at high depths after large impacts. Since the bulk of a large impactor's energy is initially delivered to the lithosphere, heat gradually flows to the mid- and lower mantle (below approximately 200 km), sustaining elevated temperatures at these depths over tens of Myr. Bombardment thus provides an extended period of time (~100 Myr; Rolf et al., 2017) over which to impart significant thermal (and by extension, chemical) heterogeneities in a planet's mantle; this is consistent with observations of Mercury (Rivera-Valentin and Barr, 2014).

The prolonged temperature effect of impactors ≳100 km in diameter is further demonstrated for the cLHB scenario in **Figure 2**, where **(a) – (b)** show the model surface area at the top of the asthenosphere (120 km depth) immediately following the largest impactor and at the simulation end (as in **Figure 1 (a)** and **(c)**). **Figure 2 (c) – (d)** shows the model surface near the CMB (400 km depth; the CMB is at 420 km) at these same times. Note the absence of any discernible heating at these depths from impactors smaller than about 100 km in diameter, contrasted with the high surface cratering in **Figure 1 (c)**.

**3.1 Crustal melting**

While in all models less than 1% of the crust is molten at any given time (**Figure 3 (a)**), the cumulative melt production as a percentage of the total crustal volume (assuming 35 km thickness and a hermeotherm in the crust of 12 °C km$^{-1}$, **Table 1**) is substantial for the 10x delivered mass cases. This arises a function of both the total mass delivered and the assumed impact velocity, and as noted earlier is a minimum estimate due to the exclusion of impactors smaller than 5 km diameter in our simulations. As a sensitivity test, we adjust the melting point by ±100 °C to consider the typical range for basaltic compositions and find that this affects instantaneous melt fractions by less than 0.1%. The amount of melt production in all models is non-negligible relative to resurfacing percentages, suggesting that both melting and the mechanical and chemical effects of bombardment played important roles in the planet's near-surface evolution during late accretion.

The resurfacing percentages for our baseline simulations are also consistent with the moderate crater saturation observed on most of Mercury's terrains (e.g., Strom and Neukum, 1988) for all but the largest craters, which may have been formed prior to late accretion (Fassett et al., 2011). Comparing these with findings for other members of the inner solar system using the same model during late accretion (Abramov et al., 2013, Abramov and Mojzsis, 2016) suggests the degree of impact-induced crustal melting on Mercury by this late bombardment far exceeded that on Earth or Mars. This is primarily a consequence of the higher average impact velocity for Mercury, which we take as 33 km s$^{-1}$, as opposed to 20 km s$^{-1}$ for Earth in Abramov et al. (2013), and 10 km s$^{-1}$ for Mars in Abramov and Mojzsis (2016).

### 3.2 Komatiite production

**Figure 3(b)** shows our rudimentary estimates for komatiite production in the asthenosphere at each timestep in each simulation. Volumes of komatiite using this approach are found to increase sharply following large impacts, but are effectively insensitive to impactors ≲100 km in diameter. At the end of the simulation, cumulative melt production is equivalent to 4% and 10% of the asthenosphere in our sLHB and cLHB scenarios, respectively. For all simulations, as expected, the amount of komatiite produced is approximately proportional to the delivered mass.

The maximum possible volume of high-Mg material in the HMR is about $2.5 \times 10^8$ km$^3$ under the assumption that it extends to the base of the crust with a regional thickness of 25 km (Padovan et al., 2015). For comparison, the cumulative volume of komatiitic melt produced in our sLHB model is $7 \times 10^8$ km$^3$, or roughly 3x the maximum possible volume of the HMR. This seems reasonable, as one would not expect all melt produced within the mantle to be deposited in the crust; most of it can be sequestered in the mantle (e.g., Roberts and Barnouin, 2012).

### 3.3 Komatiite delivery

The volume of komatiite produced in all simulations is large, and consequently is available to distribute heterogeneously through volumetrically large parts of the planet's mantle, including the source region of the HMR. Once formed at depth, plumes of komatiite-rich material in our scenario eventually reach the base of the lithosphere, where a portion of this material is sampled beneath the stagnant lid via rift volcanism as proposed by Frank et al. (2017). From this standpoint, komatiitic magmas can be delivered by lithostatic pressure-release and, closer to the surface (within 2 km or so), via decompression of gases within the magma (Wilson and Head, 1981; cf. Weider et al., 2016). The presence of gases is not an absolute requirement for eruption, however, and models based on purely lithostatic arguments exist to explain the emplacement of (for example) gas-free lunar mare basalts (Solomon, 1975; Davies and Stephenson, 1977). Mechanisms including impact-induced fracturing, uplift and mantle heating by decompression melting (Elkins-Tanton and Hager, 2005) can result from the impulse of a large impact. It has been argued that these events can alter the underlying mantle dynamics and lead to volcanism which masks the impact site (Jones et al., 2002).

As previously mentioned, in evaluating scenarios for the formation of the HMR, Frank et al. (2017) reasoned that delivery of substantial amounts of ultramafic mantle material to Mercury's surface purely by impact excavation is unlikely due to an absence of convincing geological and geophysical evidence for an impact basin; rather, the HMR is more likely the result of high-temperature volcanism. That study, however, did not rule out the possibility that volcanism could have been provoked by an impact. Our model shows that regional heating from the largest impacts is able to generate the thermal fields necessary to create conditions for the formation of komatiite-like melts on Mercury by effusive volcanism from affected mantle domains. Such a scenario is generally consistent with observations of flood volcanism by low-viscosity komatiitic lavas spread over long distances (e.g., Head et al., 2011; Byrne et al., 2013). The abundance of komatiite produced in the mantle in our simulations, offers a viable explanation for the origin of the HMR.

**4. Conclusions**

We used a 3D transient heating model for emplacement of heat into Mercury's crust and mantle by impact bombardment under two scenarios for late accretion to explain the presence of Mercury's high-magnesium region observed by MESSENGER. Our results indicate that rare, very large impactors can heat the planet's asthenosphere to temperatures at which substantial volumes of komatiite may be generated, and we suggest that this is a viable origin for high-Mg lavas on Mercury observed on Mercury's surface by MESSENGER. Plumes of komatiitic magmas produced in the asthenosphere by thermal enhancement from impact bombardments could have eventually reached the base of the crust and erupted in flood volcanism to form the HMR. We also consider the portion of Mercury's crust molten over the bombardment duration and find cumulative melt fractions at or well above 1%, pointing to their nontrivial role in resurfacing during late accretion, and the effective erasure of an earlier (pre-)Tolstojan cratering record consistent with observations (Marchi et al., 2013; Werner, 2014). In future work, we investigate the amount of resurfacing and crustal melting for all of the inner (terrestrial) planets from other sources, and for an array of probable bombardment profiles. (**5186 words**)

**Acknowledgements**

We thank Jeff Jennings (LMU-Munich) for his important contributions to this project. S.J.M. O.A., and R.B. are grateful for generous support from the Collaborative for Research in Origins (CRiO) at the University of Colorado Boulder, which is funded by the John Templeton Foundation-FfAME Origins program: The opinions expressed in this publication are those of the authors, and do not necessarily reflect the views of the John Templeton Foundation. E.A.F. was supported on a MESSENGER postdoctoral fellowship at the Carnegie Institution of Washington's Department of Terrestrial Magnetism: The MESSENGER mission was supported by the NASA Discovery Program under contracts NAS5-97271 to The Johns Hopkins University Applied Physics Laboratory and NASW-00002 to the Carnegie Institution of Washington. Continuing work on bombardment modelling by SJM is also supported by the NASA Exobiology Program (14-EXO14_2-0050; PI Simone Marchi). S.J.M. also acknowledges munificent sabbatical leave support from the Earth Life Science Institute (ELSI) at Tokyo Institute of Technology where a substantial fraction of this manuscript was generated. R.B. is supported by the Astrobiology Centre Project of the National Institutes of Natural Sciences #AB271017, the Daiwa Anglo-Japanese Foundation through a Small Grant and JSPS Kakenhi #16K17662. This work utilized the JANUS supercomputer (now retired), which was supported by the National Science Foundation (award number CNS-0821794) to the University of Colorado Boulder, the University of Colorado Denver, and the National Center for Atmospheric Research. The JANUS supercomputer and its successor, SUMMIT is operated by the University of Colorado Boulder. Discussions with S. Marchi, N. Kelly and S. Werner are gratefully acknowledged.

**Figure Captions**

**Figure 1**. Evolution of models for the classical Late Heavy Bombardment (cLHB, left column) and sawtooth (sLHB, right column) late accretion scenarios for Mercury. White isotherms are at 500 °C intervals. The scale bar at right applies to subplots (a) – (e); cratering at the surface is shown shaded in the model images. **a**) cLHB model following a strike by the largest (200 km diameter) impactor, showing the temperature effect on a local region that reaches down to the CMB (420 km depth). The immediate temperature increase in this region of the CMB is 400 °C. **b**) 10 Myr further into the bombardment simulation, showing the gradual conduction of heat upward from the CMB and the warm impact crater at the surface. **c**) Final timestep of the model, with the surface heavily cratered and the enduring effect of the largest impacts apparent in the isotherms. **d**) sLHB model following a strike by the largest (120 km diameter) impactor that induces only 200 °C heating in a local region of the CMB. **e**) Final timestep of the sLHB simulation, with moderate cratering at the surface dominated by the four largest impactors of the simulation and the lack of a strong residual temperature effect at the CMB due to the absence of impactors >120 km over the bombardment period. **f**) Temperature profiles in the cLHB scenario

for the timesteps in (a) – (c), showing ΔT of the model element at each depth through which a 200 km impactor traversed after striking the surface at 8.5 Myr into the simulation.

**Figure 2. a**) Temperature effects of bombardment in the classical Late Heavy Bombardment (cLHB). **a)** The lithosphere base (120 km) at 9 Myr, following a strike by the simulation's largest (200 km diameter) impactor at 8.5 Myr (see **Figure 1(a)**). Scale bar applies only to subplots (a) and (b). The diameter of the largest impactors is given near their impact loci. **b)** The same depth at the final timestep, showing the slow rate of cooling at earlier impact locations. **c)** The model at 400 km depth, near the core-mantle boundary (420 km), at the same timestep as in (a). Scale bar applies only to subplots (c) and (d). Immediate temperature effects of the 200 km impactor are reduced at this depth relative to 120 km depth by a factor of only ~3. **d)** The same depth at the final timestep, lacking hotspots seen in (b) that were caused by impactors too small to deliver heat to the this higher depth.

**Figure 3. a)** Percent of Mercury's crust (upper 35 km) above the melting temperature of basalt (1100 °C over time for each model, assuming a surface temperature of 120 °C and thermal gradient in the crust of 12 °C km$^{-1}$). Melt deposited in ejecta blankets is neglected. Color scheme applies to all plots: classical Late Heavy Bombardment (cLHB, red), 10x delivered mass cLHB (yellow), sawtooth Late Heavy Bombardment (sLHB, blue), and 10x delivered mass sLHB

(black). Inset: Zoom out to show full cLHB, 10x model. **b)** Komatiite production in the asthenosphere (120 – 420 km depth). Model elements at a given depth and timestep whose temperature increase from their initial values ΔT is at least 150 °C (see Sec. 2.1.5) are treated as producing their volume in komatiite. This cumulative production is shown equivalently as a percent of the total asthenosphere. **c)** Impactor distribution for each model, with the substantial majority of impactors having <50 km diameter. Because impactors of diameter ≲100 km only penetrate to shallow depths, their temperature effects in (b) are minor.

**Table captions.**

**Table 1.** Summary of target properties. A basaltic lithology (and therefore, rheology) is assumed. Latent heat of fusion, liquidus temperature, and solidus temperature are provided in the HEATING materials library (Childs, 1993); references and justifications for other values are given in the text.

**Table 2.** Statistics for late accretion bombardment scenarios on Mercury. The base of the lithosphere is taken as 120 km and the core-mantle boundary (CMB) as 420 km. The $\Delta T$ values correspond to the largest temperature increase at a single model element ($1.3 \times 10^4$ km$^2$) relative to its initial value at the indicated depth. Crustal melting calculations assume a surface temperature of 120 °C and thermal gradient in the lithosphere of 12 °C. Resurfacing is calculated based on final crater areas.

surface revealed by the MESSENGER X-Ray Spectrometer. Journal of Geophysical Research: Planets 117(E12).

Weider, S.Z., Nittler, L.R., Starr, R.D., Crapster-Pregont, E.J., Peplowski, P.N., Denevi, B.W., Head, J.W., Byrne, P.K., Hauck II, S.A., Ebel, D.S., Solomon, S.C., 2015. Evidence for geochemical terranes on Mercury: Global mapping of major elements with MESSENGER's X-Ray Spectrometer. Earth and Planetary Science Letters 416, 109–120.

Werner, S. C. (2014). Moon, Mars, Mercury: Basin formation ages and implications for the maximum surface age and the migration of gaseous planets. Earth and Planetary Science Letters 400, 54-65. doi: 10.1016/j.epsl.2014.05.019

Williams, D.A., Greeley, R., 1994. Assessment of antipodal-impact terrains on Mars. Icarus 110(2), 196–202.

Williams, D.A., Wilson, A.H., Greeley, R., 2000. A komatiite analog to potential ultramafic materials on Io. Journal of Geophysical Research: Planets 105, 1671–1684.

Williams, J.P., Ruiz, J., Rosenburg, M.A., Aharonson, O., Phillips, R.J., 2011. Insolation driven variations of Mercury's lithospheric strength. Journal of Geophysical Research: Planets 116, E01008.

Wilson, L., Head, J.W., 1981. Ascent and eruption of basaltic magma on the Earth and Moon. Journal of Geophysical Research: Solid Earth 86(B4), 2971–3001.36

| Parameter | Value |
| --- | --- |
| *Pressure derivative of the bulk modulus* | 5.5 |
| *Adiabatic bulk modulus at zero pressure* | 19.3 GPa |
| *Heat capacity* | 1200 J kg$^{-1}$ °C$^{-1}$ |
| *Density (of "anorthositic" crust)* | 2800 kg m$^{-3}$ |
| *Latent heat of fusion* | 400 kJ kg$^{-1}$ |
| *Liquidus temperature (basalt)* | 1250 °C |
| *Solidus temperature* | 1100 °C |
| *Surface temperature (at 4 Ga)* | 120 °C |
| *Crustal thickness* | 35 km |
| *Hermeothermal gradient* | 12 °C km-1, 0-120 km |
| | 0.3 °C km-1, 120-400 km |
| *Thermal conductivity* | 2.5 W m$^{-1}$ °C$^{-1}$ |

**Table 1.** Summary of target properties. A basaltic lithology (and therefore, rheology) is assumed. Latent heat of fusion, liquidus temperature, and solidus temperature are provided in the HEATING materials library (Childs, 1993); references and justifications for other values are given in the text.

| Bombardment scenario | Time [Ga] | Mass delivered [$10^{19}$ kg] | Largest impactor diameter [km] | Highest ΔT at 120 km depth [°C] | Highest ΔT, at 400 km depth [°C] | % crust melted | % resurfaced | % asthenosphere that produced komatiite |
|---|---|---|---|---|---|---|---|---|
| classical Late Heavy Bombardment (cLHB) | 3.95 - 3.85 | 3.00 | 196 | 4452 | 672 | 6 | 73 | 10 |
| 10× delivered mass cLHB | | 30.00 | 310 | 9089 | 2954 | 58 | 100 | 80 |
| sawtooth Late Heavy Bombardment (sLHB) | 4.1 - 3.7 | 0.84 | 124 | 1424 | 208 | 1 | 20 | 4 |
| 10× delivered mass sLHB | | 8.50 | 246 | 5254 | 1309 | 13 | 100 | 29 |

**Table 2.** Statistics for late accretion bombardment scenarios on Mercury. The base of the lithosphere is taken as 120 km and the core-mantle boundary (CMB) as 420 km. The ΔT values correspond to the largest temperature increase at a single model element ($\approx 1.3 \times 10^4$ km$^2$) relative to its initial value at the indicated depth. Crustal melting calculations assume a surface temperature of 120 °C and thermal gradient in the lithosphere of 12 °C. Resurfacing is calculated based on final crater areas.

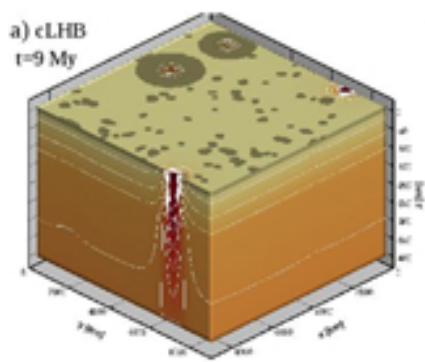
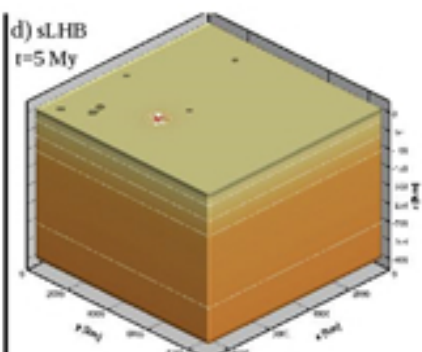
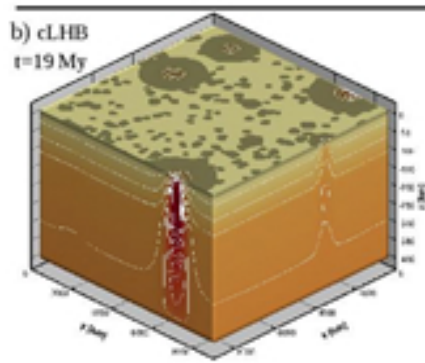
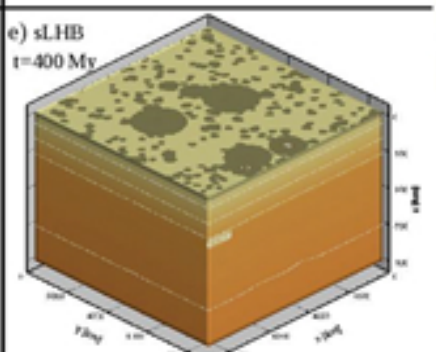
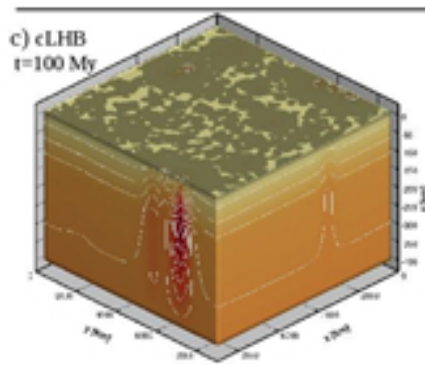
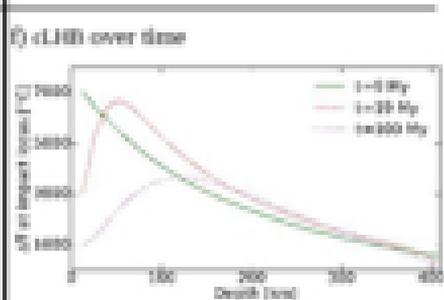
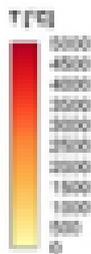

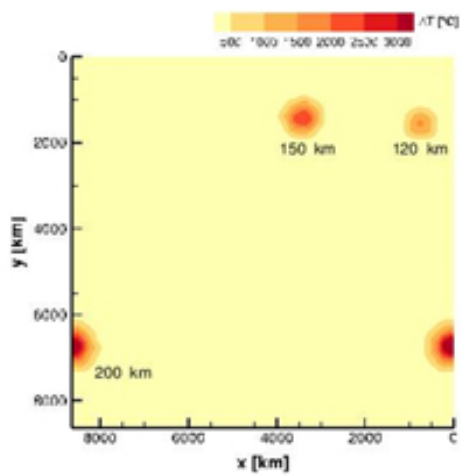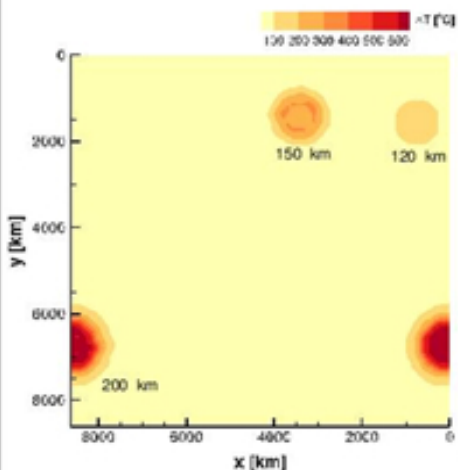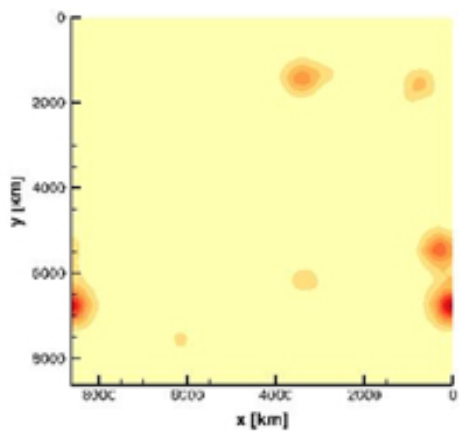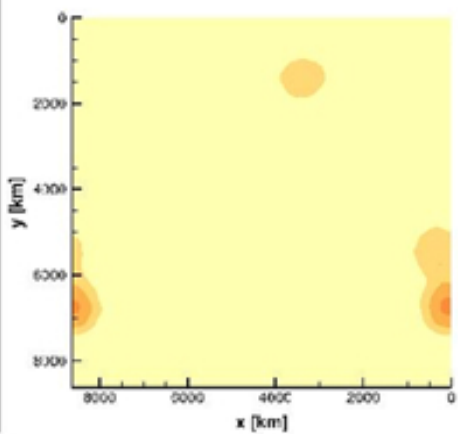

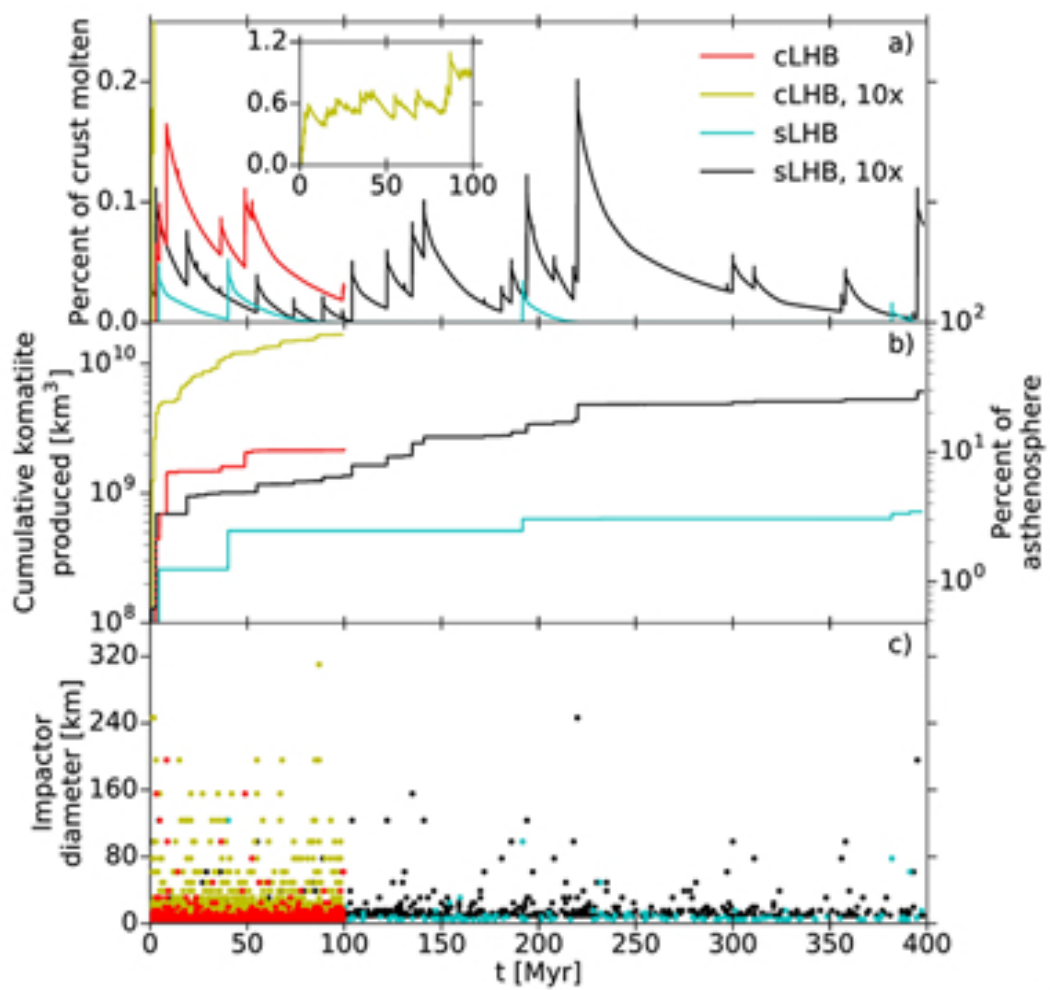